# Real-time forecasts of the 2019-nCoV epidemic in China from February 5th to February 24th, 2020

K. Roosa[1], Y. Lee[1], R. Luo[1], A. Kirpich[1], R. Rothenberg[1], J. M. Hyman[2], P. Yan[3], G. Chowell[1]

[1] Department of Population Health Sciences, School of Public Health, Georgia State University, Atlanta, GA, USA

[2] Department of Mathematics, Center for Computational Science, Tulane University, New Orleans, LA, USA

[3] Infectious Disease Prevention and Control Branch, Public Health Agency of Canada, Ottawa, Canada

**Abstract**

The initial cluster of severe pneumonia cases that triggered the 2019-nCoV epidemic was identified in Wuhan, China in December 2019. While early cases of the disease were linked to a wet market, human-to-human transmission has driven the rapid spread of the virus throughout China. The Chinese government has implemented containment strategies of city-wide lockdowns, screening at airports and train stations, and isolation of suspected patients; however, the cumulative case count keeps growing every day. The ongoing outbreak presents a challenge for modelers, as limited data are available on the early growth trajectory, and the epidemiological characteristics of the novel coronavirus are yet to be fully elucidated.

We use phenomenological models that have been validated during previous outbreaks to generate and assess short-term forecasts of the cumulative number of confirmed reported cases in Hubei province, the epicenter of the epidemic, and for the overall trajectory in China, excluding the province of Hubei. We collect daily reported cumulative case data for the 2019-nCoV outbreak for each Chinese province from the National Health Commission of China. Here, we provide 5, 10, and 15 day forecasts for five consecutive days: February 5$^{th}$ through February 9$^{th}$, with quantified uncertainty based on a generalized logistic growth model, the Richards growth model, and a sub-epidemic wave model.

Our most recent forecasts reported here are based on data up until February 9, 2020, largely agree across the three models presented and suggest an average range of 7,409 – 7,496 additional cases in Hubei and 1,128 – 1,929 additional cases in other provinces within the next five days. Models also predict an average total cumulative case count between 37,415 – 38,028 in Hubei and 11,588 – 13,499 in other provinces by February 24, 2020.

Mean estimates and uncertainty bounds for both Hubei and other provinces have remained relatively stable in the last three reporting dates (February 7th – 9th). We also observe that each of the models predicts that the epidemic has reached saturation in both Hubei and other provinces. Our findings suggest that the containment strategies implemented in China are successfully reducing transmission and that the epidemic growth has slowed in recent days.

**Introduction**

The ongoing epidemic of the novel coronavirus (2019-nCoV) is primarily affecting mainland China and can be traced back to a cluster of severe pneumonia cases identified in Wuhan, China in December 2019 (Li et al., 2020; World Health Organization, 2020). Early cases of the disease have been linked to a live animal seafood market in Wuhan, pointing to a zoonotic origin of the epidemic. However, human-to-human transmission has driven its rapid spread with a total of 37,289 confirmed cases, including 813 deaths, in China and 302 confirmed cases imported in multiple countries as of February 9, 2020 (Chinese National Health Committee). While the early transmission potential of this novel coronavirus appeared similar to that of severe acute respiratory syndrome (SARS) (Riou & Althaus, 2020), the current tally of the epidemic has already surpassed the total cases reported for the SARS outbreaks in 2002-2003 (W. World Health Organization, 2003; Wu, Leung, & Leung, 2020; Zhang et al., 2020).

The timing and location of the outbreak facilitated the rapid transmission of the virus within a highly mobile population. The initial reporting of observed cases occurred during the traditional

Chinese New Year, when the largest population movement in China takes place every year (Ai et al., 2020). Further, Wuhan is a highly populated city with more than 11 million residents and is connected to many cities in China through public transportation, such as buses, trains, and flights (Lai et al., 2020; Read, Bridgen, Cummings, Ho, & Jewell, 2020). In the absence of pharmaceutical interventions, rapid action was required by the Chinese government to mitigate transmission within and outside of Wuhan.

On January 23, 2020, the Chinese government implemented a strict lockdown of Wuhan, followed by several nearby cities in subsequent days; the lockdown includes temporarily restricting public transportation and advising residents to remain at home (Du et al., 2020; Wu et al., 2020). Further, many high-speed rail stations and airports have implemented screening measures to detect travelers with a fever, specifically those traveling from Wuhan, and those with a fever are referred to public hospitals (Lai et al., 2020; Wu et al., 2020). Within hospitals, patients who fulfill clinical and epidemiological characteristics of 2019-nCoV are immediately isolated.

The number of 2019-nCoV cases in Wuhan quickly outnumbered the available number of beds in hospitals, putting a substantial burden on the healthcare system. Consequently, the government rapidly built and launched two new hospitals with capacity for 1,600 and 1,000 beds, respectively, in Wuhan in addition to the existing 132 quarantine sites with more than 12,500 beds (Steinbuch, 2020). To anticipate additional resources to combat the epidemic, mathematical and statistical modeling tools can be useful to generate timely short-term forecasts of reported cases. These predictions can include estimates of expected morbidity burden that can help guide public health officials preparing the medical care and other resources needed to confront the epidemic. Short-term forecasts can also guide the intensity and type of interventions needed to mitigate an epidemic (Funk, Camacho, Kucharski, Eggo, & Edmunds, 2016; Shanafelt, Jones, Lima, Perrings, & Chowell, 2017). In the absence of vaccines or antiviral drugs for 2019-nCoV, the effective implementation of nonpharmaceutical interventions, such as personal protection and social distancing, will be critical to bring the epidemic under control.

In this emerging epidemic, the epidemiological data is limited, and the epidemiological parameters needed to calibrate elaborate mechanistic transmission models are not yet fully elucidated. Real-time short-term forecasts must be based on dynamic phenomenological models that have been validated during previous outbreaks (Chowell et al., 2016; Pell, Kuang, Viboud, & Chowell, 2018). We employ several dynamic models to generate and assess 5, 10, and 15 day ahead forecasts of the cumulative number of confirmed cases in Hubei province, the epicenter of the epidemic, and the overall trajectory of the epidemic in China excluding the province of Hubei.

**Methods**

*Data*

We obtained daily updates of the cumulative number of reported confirmed cases for the 2019-nCoV epidemic across provinces in China from the National Health Commission of China website (Chinese National Health Commission). The data contains 34 areas, including provinces, municipalities, autonomous regions, and special administrative regions; here we refer to the regions collectively as provinces. Data updates were collected daily at 12 pm (GMT-5), between January 22, 2020 and February 9, 2020. The short time-series is affected by irregularities and reporting lags, so the cumulative curves are more stable and likely yield more stable and reliable estimates. Therefore, we analyze the cumulative trajectory of the epidemic in Hubei province, the epicenter of the outbreak, as well as the cumulative aggregate trajectory of all other provinces.

*Models*

We generate short-term forecasts in real-time using three phenomenological models that have been previously used to derive short-term forecasts for a number of epidemics for several infectious diseases, including SARS, Ebola, pandemic influenza, and dengue (Chowell, Tariq, & Hyman, 2019; Pell et al., 2018; Wang, Wu, & Yang, 2012). The generalized logistic growth model (GLM) extends the simple logistic growth model to accommodate sub-exponential growth dynamics with a *scaling of growth* parameter, $p$ (Viboud, Simonsen, & Chowell, 2016). The Richards model also includes a scaling parameter, $a$, to allow for deviation from the symmetric logistic curve (Chowell, 2017; Richards, 1959; Wang et al., 2012). We also include a recently developed sub-epidemic wave model that supports complex epidemic trajectories, including multiple peaks (i.e., SARS in Singapore (Chowell et al., 2019)). In this approach, the observed reported curve is assumed to be the aggregate of multiple underlying sub-epidemics (Chowell et al., 2019). A detailed description for each of the models is included in the Supplement.

*Short-term forecasts*

We calibrate each model to the daily cumulative reported case counts for Hubei and other provinces (all except Hubei). While the outbreak began in December 2019, available data on cumulative case counts were available starting on January 22, 2020. Therefore, the first calibration process includes 15 observations: from January 22, 2020 to February 5, 2020. Each subsequent calibration period increases by one day with each new published daily data point, with the last calibration period comprising the period from January 22, 2020 to February 9, 2020 (19 data points).

We estimate the best-fit model solution to the reported data using nonlinear least squares fitting. This process yields the set of model parameters $\Theta$ that minimizes the sum of squared errors between the model $f(t, \Theta)$ and the data $y_t$; where $\Theta_{GLM} = (r, p, K)$, $\Theta_{Rich} = (r, a, K)$, and $\Theta_{Sub} = (r, p, K_0, q, C_{thr})$ correspond to the estimated parameter sets for the GLM, the Richards model, and the sub-epidemic model, respectively; parameter descriptions are provided in the Supplement. Thus, the best-fit solution $f(t, \widehat{\Theta})$ is defined by the parameter set $\widehat{\Theta} = arg\ min \sum_{t=1}^{n}(f(t, \Theta) - y_t)^2$. We fix the initial condition to the first data point.

We then use a parametric bootstrap approach to quantify uncertainty around the best-fit solution, assuming a Poisson error structure. A detailed description of this method is provided in prior studies (Chowell, 2017; Roosa & Chowell, 2019). The models are refitted to the $M = 200$ bootstrap datasets to obtain $M$ parameter sets, which are used to define 95% confidence intervals for each parameter. Each of the $M$ model solutions to the bootstrap curves is used to generate $m = 30$ simulations extended through a forecasting period of 15 days. These 6,000 ($M \times m$) curves are used to construct the 95% prediction intervals for the forecasts.

**Results**

We generated 5, 10, and 15 day ahead forecasts for Hubei and other provinces excluding Hubei for 5 consecutive starting dates: February 5, 2020 to February 9, 2020. Figures 1 – 3 represent the range of 5, 10, and 15 day ahead forecasts, respectively, by the date generated, and we compare the daily short-term forecasts of cumulative case counts across dates as more data become available. Current cumulative reported case counts as of February 9, 2020 are 27,100 for Hubei and 10,189 in other provinces (Chinese National Health Commission).

*Model calibration*

Our results for Hubei province indicate that the parameter estimates for the three models tend to stabilize and decrease in uncertainty as more data become available (Supplemental Table 1). In particular, the growth rate $r$ decreases and appears to be converging over time, particularly for the GLM and sub-epidemic model. Parameter $K$ also follows this general trend, with prediction intervals decreasing significantly in width as more data become available. Importantly, the $p$ estimates from the GLM indicate that the epidemic growth in Hubei is close to exponential ($p$ =0.99 (95%CI: 0.98, 1) – February 9th). Further, growth rate and scaling parameter estimates have remained relatively stable over the last three reporting dates, while estimates of $K$ are still declining. This may correlate with the effectiveness of control measures or the slowing of the epidemic.

For the trajectory that aggregates all other provinces (excluding Hubei), the parameter estimates follow trends that differ from those for Hubei (Supplemental Table 2). While the three models estimated stable and nearly equivalent growth rates in Hubei, the estimated growth rates for other provinces vary across models and do not follow a distinct trend as more data become available. However, the scaling and size parameters remain relatively stable across all dates. Further, the $p$ estimates from the GLM reveal a consistent sub-exponential growth pattern in other provinces ($p$ = 0.67 (95%CI: 0.64, 0.70) – February 9th).

*5 days ahead*

The latest 5 day ahead forecasts, generated on February 9, 2020, estimate an average of 34,509 – 34,596 total cumulative cases in Hubei by February 14, 2020 across the three models (Figure 1a). For other provinces, the models predict an average range of 11,317 – 12,118 cumulative cases by February 14 (Figure 1b). Based on cumulative reported cases as of February 9th, these estimates correspond with an average of 7,409 – 7,496 additional cases in Hubei and 1,128 – 1,929 additional cases in other provinces within the next 5 days.

Comparing the 5-day ahead forecasts generated daily on February 5 – 9, 2020, the GLM and Richards models yield comparable prediction intervals in Hubei, while the sub-epidemic model yields wider intervals than the other models. Also, 5 day ahead forecasts from the sub-epidemic model on February 5th and 6th predict significantly higher case counts in Hubei compared to forecasts generated on February 7th and beyond (Figure 1a). For other provinces, the GLM and Richards model yield intervals of similar widths, but the GLM predicts higher case counts than the Richards model across all dates (Figure 1b). Further, the sub-epidemic model has significantly wider prediction intervals compared to the other models for all forecasts for other provinces. While the uncertainty of the predictions decreases as more data became available in Hubei, the uncertainty of the predictions for other provinces remain relatively stable, compared to forecasts from earlier dates.

*10 days ahead*

The 10 day ahead forecasts generated on February 9, 2020 from the three models estimate between 36,854 – 37,230 cumulative cases, on average, in Hubei by February 19, 2020 (Figure 2a). For other provinces, the latest 10 day ahead forecasts predict average cumulative case counts between 11,549 – 13,069 cases across the three models (Figure 2b). These estimates correspond with an additional 9,754 – 10,130 cases in Hubei and an additional 1,360 – 2,880 cases reported in other provinces on average in the next 10 days.

10 day ahead forecasts of case counts in Hubei generated on February 5$^{th}$ show significantly different results between the GLM and Richards versus the sub-epidemic model, with the sub-epidemic model predicting significantly higher case counts (Figure 2a). For forecasts generated after February 5$^{th}$, the prediction intervals of the three models are comparable, with the GLM intervals having the lowest uncertainty, followed by the Richards model (Figure 2a). For other provinces, the sub-epidemic model yields significantly wider prediction intervals than the other two models. Like the 5 day ahead forecasts, the 10 day ahead prediction intervals become increasingly narrow for Hubei when including more data, but uncertainty remains relatively stable in other provinces.

*15 days ahead*

The latest 15 day ahead forecasts predict a cumulative reported case count between 37,415 – 38,028 cases, on average, in Hubei by February 24, 2020. Further, the latest 15 day ahead forecasts suggest an average cumulative case count between 11,588 – 13,499 cases for other provinces. These forecasts correspond with an additional 10,315 – 10,928 cases in Hubei and an additional 1,399 – 3,310 cases in other provinces within the next 15 days.

Again, the sub-epidemic model yields significantly higher forecasts for Hubei on February 5$^{th}$, compared to the other models and compared to subsequent prediction intervals on following dates (Figure 3a). The width of prediction intervals decreases as more data are included for each of the models in both Hubei and other provinces. This is consistent with shorter-term forecasts in Hubei but differs from the pattern of shorter-term forecasts in other provinces.

**Discussion**

In this report, we provide timely short-term forecasts of the cumulative number of reported cases of the 2019-nCoV epidemic in Hubei province and other provinces in China as of February 9, 2020. As the epidemic continues, we are also publishing online daily 10-day ahead forecasts including each of the models presented here (Roosa & Chowell, 2020). Based on the three models calibrated to data up until February 9, 2020, we forecast a cumulative number of reported cases between 37,415 – 38,028 in Hubei Province and 11,588 – 13,499 in other provinces by February 24, 2020.

Our models yield a good visual fit to the epidemic curves, based on residuals, with the sub-epidemic model outperforming the other models in terms of the mean squared error (MSE) (Supplemental Tables 1 & 2). Parameter estimation results from the GLM consistently show that the epidemic growth is near exponential in Hubei and sub-exponential in other provinces.

Overall, models predict similar ranges of short-term forecasts, except for those generated on February 5$^{th}$, where the sub-epidemic model predicts significantly higher case counts than the other two models (Figures 1 – 3). The sub-epidemic model predicts similar ranges to the other models for subsequent dates, so the higher ranges on February 5th may indicate that more data are required to inform the parameters of the sub-epidemic model.

We observe that the width of the prediction intervals decreases on average as more data are included for forecasts in Hubei; however, this pattern is not obvious for our analysis based on other provinces. This can, in part, be attributed to the smaller case counts and smaller initial prediction interval range seen in other provinces. Mean predictions and associated uncertainty remain relatively stable in other provinces though, while the mean estimates of 10 and 15 days ahead decrease significantly in Hubei (Figures 2 & 3). This suggests that the epidemic will last longer in Hubei compared to other provinces (Figures 4 – 6), which may be attributed to intensive control efforts and large-scale social distancing interventions. Therefore, it is not necessarily surprising that estimates from earlier dates, specifically prior to saturation, yield predictions with higher uncertainty.

We retrieve the data from the Chinese media conglomerate Tencent (Chinese National Health Commission); however, the data show small differences in case counts compared to data of the epidemic reported by other sources (Johns Hopkins University Center for Systems Science and Engineering, 2020). Importantly, the curves of confirmed cases that we employ in our study are reported according to reporting date and could be influenced by testing capacity and other related factors. Further, there may be significant delays in identifying, isolating, and reporting cases in Hubei due to the magnitude of the epidemic, which could influence our predictions. Incidence curves according to the date of symptom onset could provide a clearer picture of the transmission dynamics during an epidemic. We also note that we analyzed the epidemic curves starting on

January 22, 2020, but the epidemic started in December 2019. Hence, the first data point accumulates cases up until January 22, 2020, as data were not available prior to this date.

The 2019-nCoV outbreak in China poses a major challenge for modelers, as there are limited data available on the early growth trajectory, and epidemiological characteristics of the novel strain have not been fully elucidated. Our timely short-term forecasts based on phenomenological models can be useful for real-time preparedness, such as anticipating the required number of hospital beds and other medical resources, as they provide an estimate of the number of cases hospitals will need to prepare for in the coming days. In future work, we plan to report the results of a retrospective analysis of forecasting performance across models based on various performance metrics.

In conclusion, our most recent forecasts, based on data for the last three days (February $7^{th}$ – $9^{th}$), remained relatively stable. These models predict that the epidemic has reached a saturation point for both Hubei and other provinces. This likely reflects the impact of the wide spectrum of social distancing measures implemented by the Chinese government, which likely helped stabilize the epidemic. The forecasts presented are based on the assumption that current mitigation efforts will continue.

*Acknowledgements*

We thank Homma Rafi (Director of Communications, School of Public Health, Georgia State University) for creating and maintaining the online record of daily short-term forecasts.

*Funding*

GC is supported by NSF grants 1610429 and 1633381.

*Ethics*

Not applicable.

*Data, code and materials*

Data will be made available in an online repository upon acceptance of manuscript.

*Competing Interests*

Authors declare no competing interests.

*Author Contributions*

KR and GC conducted forecasts and data analysis; YL retrieved and managed data; All authors contributed to writing and revising subsequent versions of the manuscript. All authors read and approved the final manuscript.

**Figure 1.** Forecasting results for 5-days ahead estimates, generated daily from February 5 – 9, 2020, of cumulative reported cases in Hubei (a) and other provinces (b). The mean case estimate is represented by the dots, while the lines represent the 95% prediction intervals for each model.

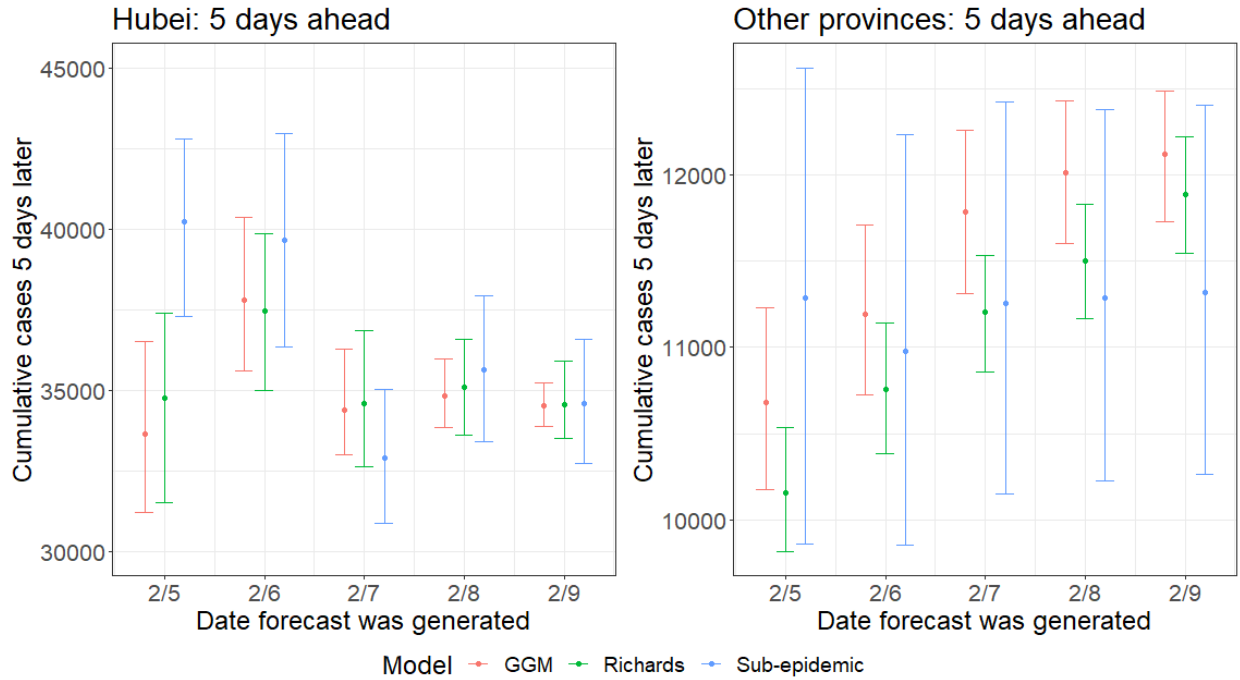

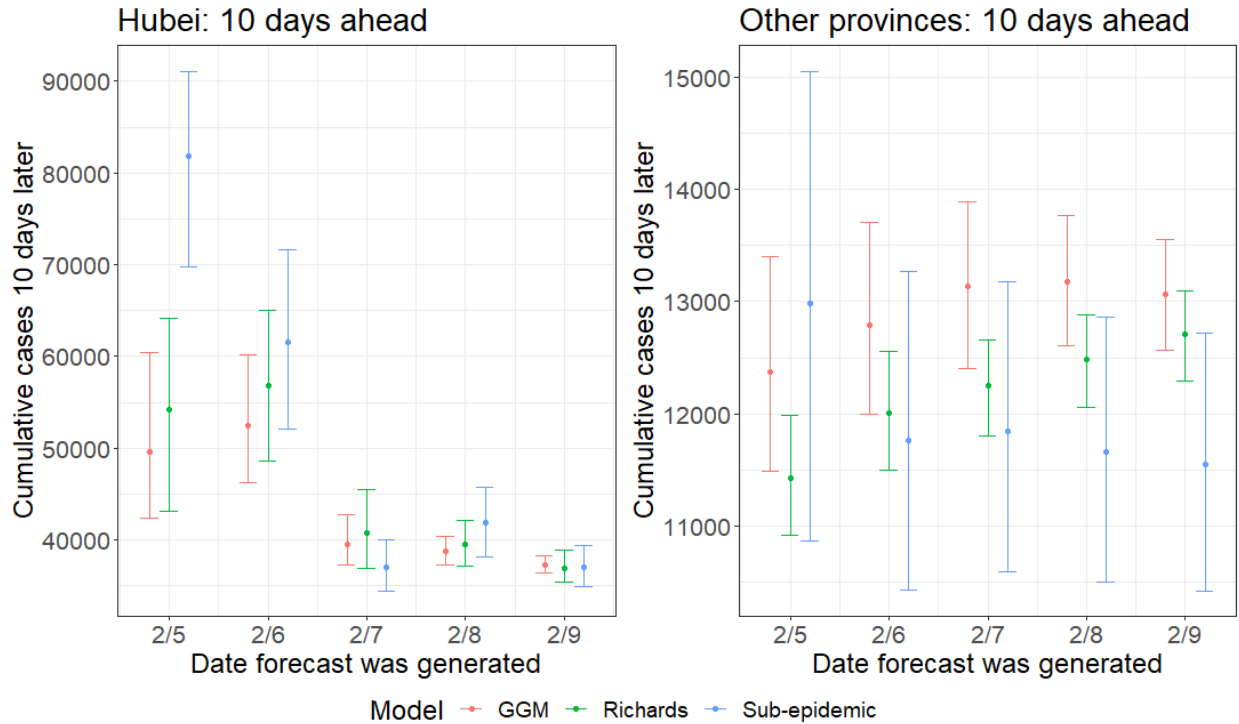

**Figure 2.** Forecasting results for 10-days ahead estimates, generated daily from February 5 – 9, 2020, of cumulative reported cases in Hubei (a) and other provinces (b). The mean case estimate is represented by the dots, while the lines represent the 95% prediction intervals for each model.

**Figure 3.** Forecasting results for 15-days ahead estimates, generated daily from February 5 – 9, 2020, of cumulative reported cases in Hubei (a) and other provinces (b). The mean case estimate is represented by the dots, while the lines represent the 95% prediction intervals for each model.

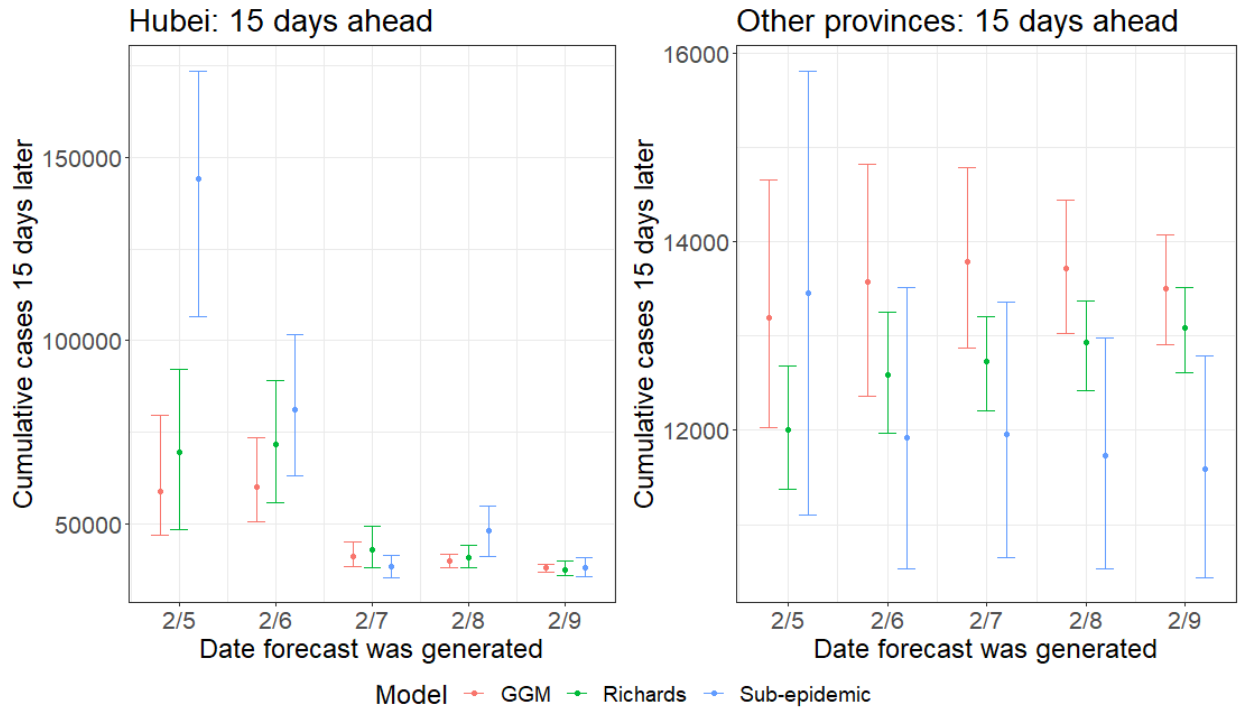

**Figure 4.** 15-day ahead GLM forecasts of cumulative reported 2019-nCoV cases in China – Hubei and other provinces – generated on February 9, 2020.

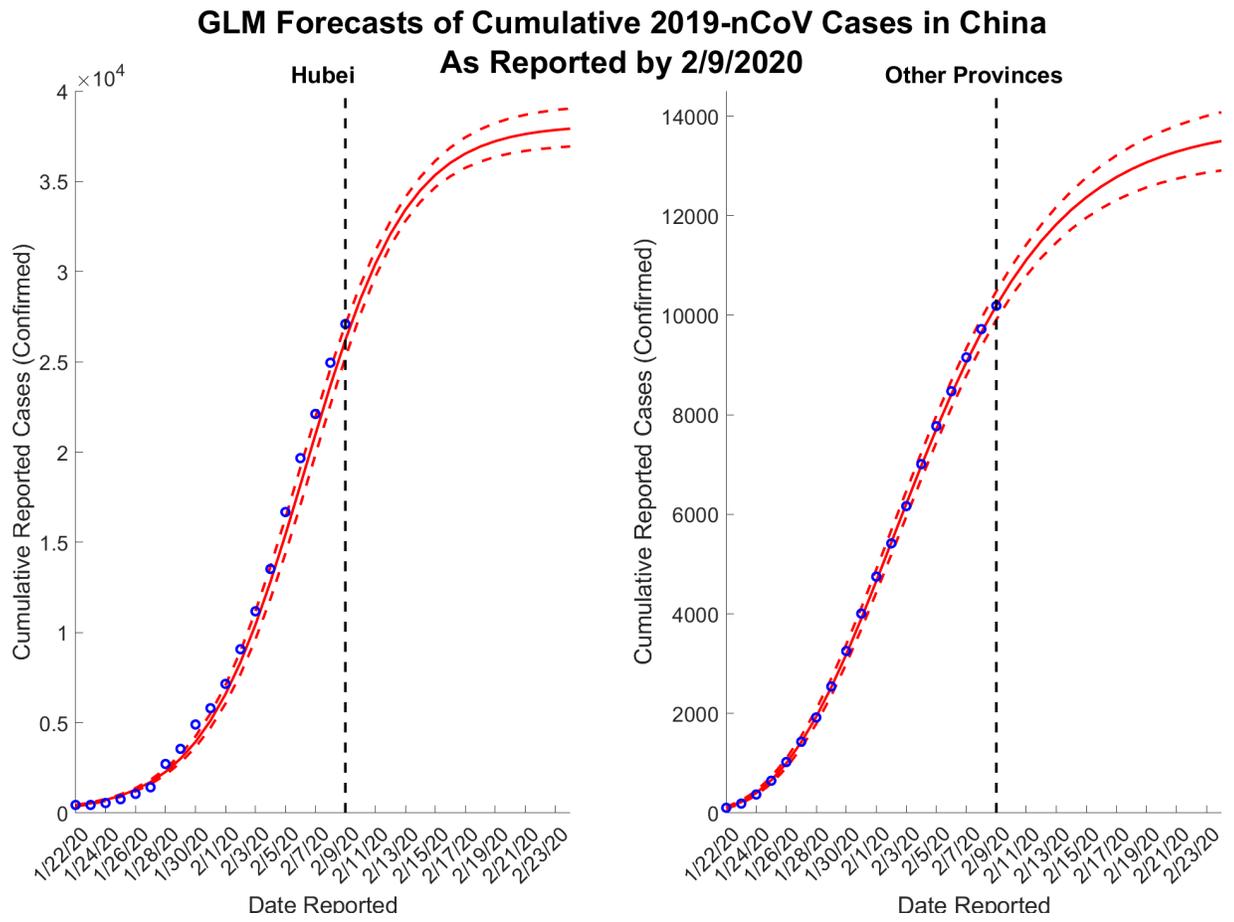

**Figure 5.** 15-day ahead Richards forecasts of cumulative reported 2019-nCoV cases in China – Hubei and other provinces – generated on February 9, 2020.

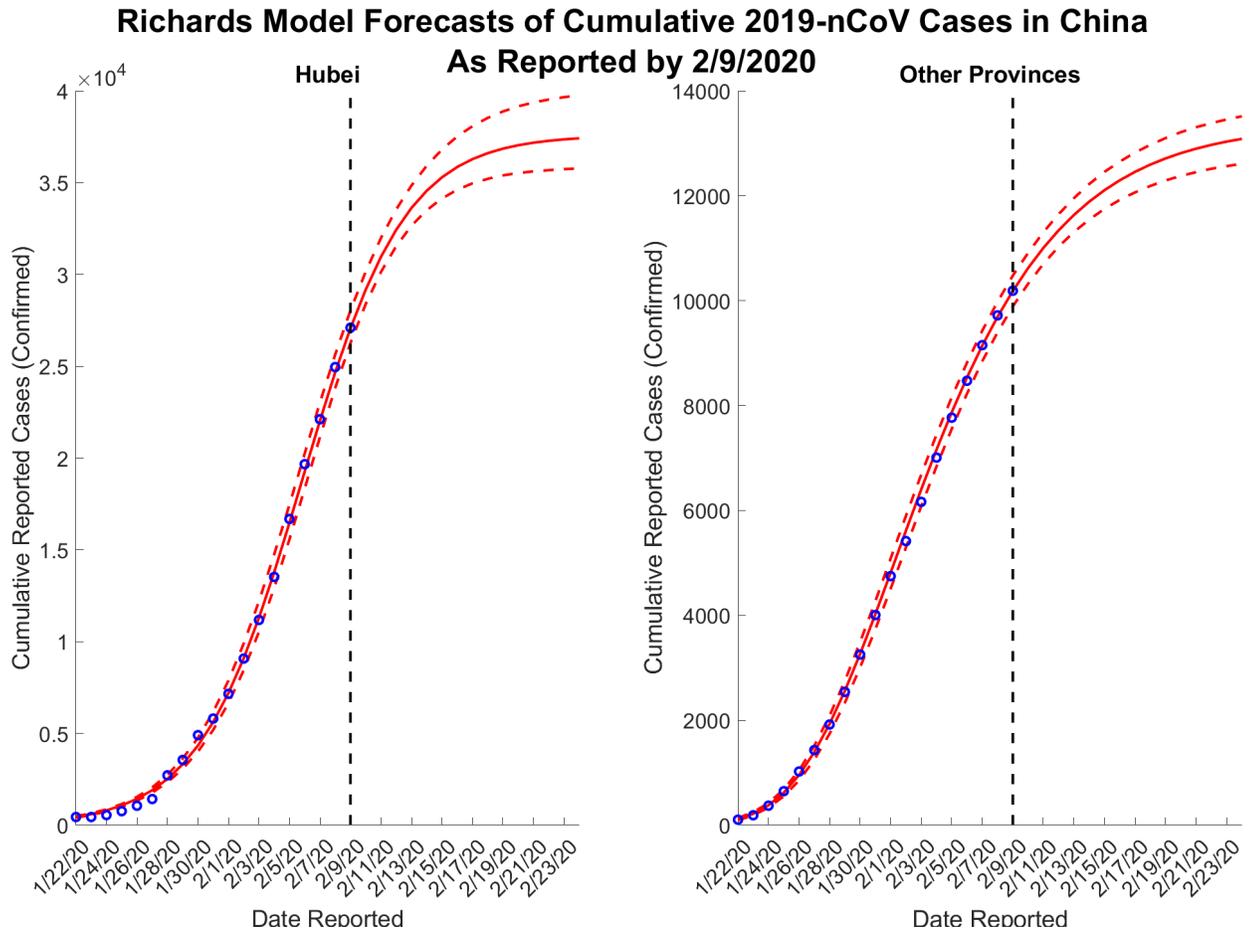

**Figure 6.** 15-day ahead sub-epidemic model forecasts of cumulative reported 2019-nCoV cases in China – Hubei and other provinces – generated on February 9, 2020.

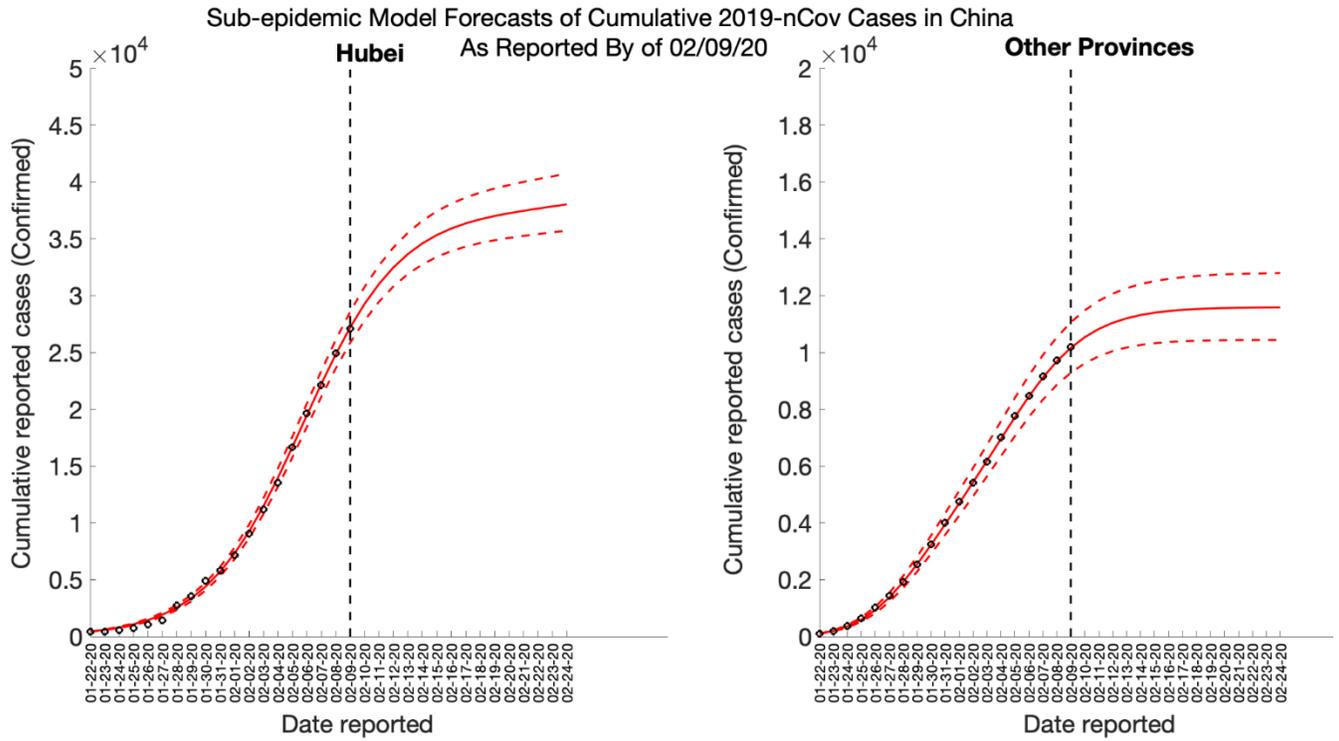

# Appendix

**Model descriptions**

*GLM*

The generalized logistic growth model (GLM) is an extension of the simple logistic growth model that includes an additional parameter, *p*, to allow for *scaling of growth*; *p* = 1 indicates early exponential growth, *p* = 0 represents constant growth, and 0 < *p* < 1 accommodates early sub-exponential or polynomial growth (G. Chowell et al., 2020; Ganyani, Roosa, Faes, Hens, & Chowell, 2018; Roosa, Luo, & Chowell, 2019; Viboud, Simonsen, & Chowell, 2016). The GLM is defined by the differential equation:

$$C'(t) = rC(t)^p \left(1 - \frac{C(t)}{K}\right)$$

where *C(t)* represents the cumulative number of cases at time *t*, *r* is the growth rate, *p* is the scaling of growth parameter, and *K* is the carrying capacity or final epidemic size.

Richards model

The Richards model is a 3-parameter extension of the simple logistic growth model that includes a scaling parameter, *a* (Gerardo Chowell, 2017; Richards, 1959; Wang, Wu, & Yang, 2012). The Richards model is defined by the differential equation:

$$C'(t) = rC(t)\left(1 - \left(\frac{C(t)}{K}\right)^a\right)$$

where $C(t)$ represents the cumulative number of cases at time $t$, $r$ is the growth rate, $K$ is the final epidemic size, and the exponent $a$ measures the deviation from the symmetric s-shaped dynamics of the simple logistic curve.

*Sub-epidemic model*

The most flexible extension employed here is the recently developed sub-epidemic wave model, which supports complex epidemic trajectories by assuming the reported aggregate curve is shaped by multiple underlying sub-epidemics (G. Chowell, Tariq, & Hyman, 2019). For this approach, each sub-epidemic is modeled using the GLM, where the growth rate $r$ and scaling parameter $p$ are the same across sub-epidemics. We model an epidemic wave composed of $n$ overlapping sub-epidemics as follows:

$$C_i'(t) = rA_{i-1}(t)C_i(t)^p \left(1 - \frac{C_i(t)}{K_i}\right),$$

where $C_i(t)$ is the cumulative number of infections for sub-epidemic $i$, and $K_i$ is the size of the $i^{th}$ sub-epidemic ($i = 1, …, n$) (G. Chowell et al., 2019). Thus, when $n = 1$, the model returns to the single-equation GLM, as presented above.

We model the timing of onset for each subsequent wave with a regular structure, such that the $(i+1)^{th}$ sub-epidemic begins when $C_i(t)$ exceeds the threshold $C_{thr}$, and the $(i+1)^{th}$ sub-epidemic begins before the $i^{th}$ sub-epidemic is complete. We model the size of consecutive sub-epidemics $(K_i)$ such that the size declines exponentially. Thus,

$$K_i = K_0 e^{-q(i-1)},$$

where $K_0$ is the size of the first sub-epidemic ($K_1 = K_0$), and $q$ is the rate of decline of consecutive sub-epidemics, where $q = 0$ indicates no decline. Further, the total final size of the epidemic is given by:

$$K_{tot} = \sum_{i=1}^{n_{tot}} K_0 e^{-q(i-1)} = \frac{K_0(1 - e^{-q n_{tot}})}{1 - e^{-q}}$$

where $n_{tot}$ is the finite number of overlapping sub-epidemics and is calculated as

$$n_{tot} = \left\lfloor -\frac{1}{q} \ln\left(\frac{C_{thr}}{K_0}\right) + 1 \right\rfloor$$

| GLM | r (95% CI) | p (95% CI) | K (95% CI) | MSE |
|---|---|---|---|---|
| 2/5/2020 | 0.54 (0.39, 0.72) | 0.92 (0.88, 0.96) | 68,730 (48,730, 104,110) | 72,787 |
| 2/6/2020 | 0.56 (0.40, 0.72) | 0.91 (0.88, 0.95) | 67,484 (52,358, 91,421) | 68,967 |
| 2/7/2020 | 0.36 (0.29, 0.43) | 0.97 (0.95, 1) | 41,700 (39,001, 44,606) | 72,918 |
| 2/8/2020 | 0.34 (0.29, 0.43) | 0.98 (0.95, 1) | 40,253 (38,528, 42,232) | 69,741 |
| 2/9/2020 | 0.31 (0.29, 0.36) | 0.99 (0.98, 1) | 38,134 (37,198, 39,251) | 69,593 |
| **Richards** | r (95% CI) | a (95% CI) | K (95% CI) | MSE |
| 2/5/2020 | 0.34 (0.30, 0.41) | 0.49 (0.28, 0.80) | 82,180 (50,940, 139,290) | 70,230 |
| 2/6/2020 | 0.35 (0.30, 0.41) | 0.46 (0.29, 0.73) | 86,990 (54,280, 72,990) | 66,528 |
| 2/7/2020 | 0.31 (0.29, 0.34) | 0.83 (0.61, 1.03) | 43,616 (38,875, 50,460) | 72,766 |
| 2/8/2020 | 0.30 (0.29, 0.32) | 0.87 (0.73, 1.05) | 41,427 (38,239, 44,887) | 69,775 |
| 2/9/2020 | 0.30 (0.28, 0.32) | 1.01 (0.84, 1.15) | 37,585 (35,974, 40,128) | 69,588 |
| **Sub-epidemic** | r (95% CI) | p (95% CI) | $K_0$ (95% CI) | MSE |
| 2/5/2020 | 0.98 (0.90, 1.1) | 0.84 (0.82, 0.85) | 429,000 (168,000, 943,000) | 55,491 |
| 2/6/2020 | 0.78 (0.62, 0.98) | 0.87 (0.84, 0.90) | 85,500 (63,300, 122,000) | 57,640 |
| 2/7/2020 | 0.32 (0.30, 0.36) | 0.99 (0.98, 1) | 38,200 (36,700, 40,100) | 70,756 |
| 2/8/2020 | 0.39 (0.33, 0.46) | 0.97 (0.95, 0.99) | 41,500 (39,300, 43,700) | 68,610 |
| 2/9/2020 | 0.31 (0.30, 0.34) | 1.0 (0.98, 1) | 37,500 (36,800, 38,600) | 68,751 |

**Supplemental Table 1.** Hubei: GLM, Richards, and sub-epidemic model parameter estimates and mean squared error (MSE) for the best-fit solution. Best-fit parameter estimates are presented with the 95% confidence intervals obtained from the *M* bootstrap solutions.

**Supplemental Table 2.** Other Provinces: GLM, Richards, and sub-epidemic model parameter estimates and mean squared error (MSE) for the best-fit solution. Best-fit parameter estimates are

| GLM | r (95% CI) | p (95% CI) | K (95% CI) | MSE |
|---|---|---|---|---|
| 2/5/2020 | 4.08 (2.96, 5.32) | 0.67 (0.63, 0.71) | 13,734 (12,420, 15,351) | 1,695 |
| 2/6/2020 | 4.21 (3.21, 5.47) | 0.66 (0.63, 0.70) | 14,129 (13,034, 15,663) | 1,673 |
| 2/7/2020 | 4.30 (3.24, 5.51) | 0.66 (0.63, 0.70) | 14,332 (13,345, 15,437) | 1,632 |
| 2/8/2020 | 4.29 (3.27, 5.27) | 0.66 (0.63, 0.70) | 14,191 (13,294, 15,103) | 1,590 |
| 2/9/2020 | 4.05 (3.36, 4.98) | 0.67 (0.64, 0.70) | 13,826 (13,171, 14,481) | 1,776 |

| Richards | r (95% CI) | a (95% CI) | K (95% CI) | MSE |
|---|---|---|---|---|
| 2/5/2020 | 2.18 (1.15, 2.98) | 0.08 (0.06, 0.17) | 12,379 (11,589, 13,020) | 3,660 |
| 2/6/2020 | 2.49 (1.66, 3.0) | 0.07 (0.06, 0.11) | 12,993 (12,391, 13,584) | 4,178 |
| 2/7/2020 | 2.14 (1.24, 2.90) | 0.08 (0.06, 0.15) | 13,106 (12,555, 13,585) | 5,593 |
| 2/8/2020 | 3.02 (2.52, 3.41) | 0.06 (0.05, 0.06) | 13,573 (13,072, 14,092) | 4,547 |
| 2/9/2020 | 2.25 (1.15, 3.0) | 0.08 (0.06, 0.16) | 13,374 (12,862, 13,808) | 5,305 |

| Sub-epidemic | r (95% CI) | p (95% CI) | $K_0$ (95% CI) | MSE |
|---|---|---|---|---|
| 2/5/2020 | 2.1 (1.8, 2.6) | 0.78 (0.75, 0.81) | 7,240 (6,750, 7,760) | 1,380 |
| 2/6/2020 | 1.9 (1.6, 2.4) | 0.80 (0.76, 0.83) | 6,420 (5,950, 6,970) | 1,412 |
| 2/7/2020 | 1.9 (1.6, 2.4) | 0.80 (0.76, 0.83) | 6,420 (5,930, 6,930) | 1,333 |
| 2/8/2020 | 1.8 (1.5, 2.1) | 0.81 (0.79, 0.83) | 6,070 (5,750, 6,580) | 1,278 |
| 2/9/2020 | 1.6 (1.4, 1.8) | 0.83 (0.81, 0.85) | 5,900 (5,680, 6,310) | 1,215 |

presented with the 95% confidence intervals obtained from the $M$ bootstrap solutions.